\documentclass[fleqn,usenatbib]{mnras}

\usepackage{mathptmx}

\usepackage[T1]{fontenc}
\usepackage{ae,aecompl}

\usepackage{graphicx}	
\usepackage{amsmath}	
\usepackage{amssymb}	
\usepackage{upgreek}    

\newcommand{\Msol}{\,M$_{\sun}$}
\newcommand{\Meraxes}{\textsc{Meraxes}}
\newcommand{\Tiamat}{\textit{Tiamat}}
\newcommand{\zelv}{$z{\sim}$11}

\title[GN-z11 analogues in DRAGONS]{Dark-ages reionization and galaxy-formation
  simulation -- VI.\@ The origins and fate of the highest known redshift galaxy}

\author[S. J. Mutch et al.]{%
Simon J. Mutch,$^{1}$\thanks{E-mail: smutch@unimelb.edu.au (SJM)}
Chuanwu Liu,$^{1}$
Gregory B. Poole,$^{1}$
Paul M. Geil,$^{1}$
Alan R. Duffy,$^{2,1}$\newauthor%
Michele Trenti,$^{1}$
Pascal A. Oesch,$^{3}$
Garth D. Illingworth$^{4}$
Andrei Mesinger,$^{5}$\newauthor%
J. Stuart B. Wyithe,$^{1}$
\\
$^{1}$School of Physics, University of Melbourne, Parkville, VIC 3010,
Australia\\
$^{2}$Centre for Astrophysics and Supercomputing, Swinburne University of
Technology, PO Box 218, Hawthorn, VIC 3122, Australia\\
$^{3}$Yale Center for Astronomy and Astrophysics, Yale University, New Haven,
CT 06511, USA\\
$^{4}$UCO/Lick Observatory, University of California, Santa Cruz, 1156 High St,
Santa Cruz, CA 95064, USA\\
$^{5}$Scuola Normale Superiore, Piazza dei Cavalieri 7, I-56126 Pisa, Italy
}

\date{Accepted 2016 August 29\@. Received 2016 August 29;\@ in original form 2016 May 24}

\pubyear{2016}

\begin{document}
\label{firstpage}
\pagerange{\pageref{firstpage}--\pageref{lastpage}}
\maketitle

\begin{abstract}
  Using Hubble data, including new grism spectra, Oesch et al.  recently
  identified GN-z11, an $M_\textrm{UV}$=-21.1 galaxy at $z$=11.1 (just 400\,Myr
  after the big bang). With an estimated stellar mass of
  $\sim$10$^9$\,M$_{\sun}$, this galaxy is surprisingly bright and massive,
  raising questions as to how such an extreme object could form so early in the
  Universe.  Using \Meraxes{}, a semi-analytic galaxy-formation model developed
  as part of the Dark-ages Reionization And Galaxy-formation Observables from
  Numerical Simulations (DRAGONS) programme, we investigate the potential
  formation mechanisms and eventual fate of GN-z11.  The volume of our
  simulation is comparable to that of the discovery observations and possesses
  two analogue galaxies of similar luminosity to this remarkably bright system.
  Existing in the two most massive subhaloes at $z$=11.1
  ($M_\textrm{vir}$=1.4$\times 10^{11}$\,M$_{\sun}$ and
  6.7$\times 10^{10}$\,M$_{\sun}$), our model analogues show excellent agreement
  with all available observationally derived properties of GN-z11.  Although
  they are relatively rare outliers from the full galaxy population at
  high-$z$, they are no longer the most massive or brightest systems by $z$=5.
  Furthermore, we find that both objects possess relatively smooth, but
  extremely rapid mass growth histories with consistently high star formation
  rates and UV luminosities at $z{>}11$, indicating that their brightness is
  not a transient, merger-driven feature.  Our model results suggest that
  future wide-field surveys with the \textit{James Webb Space Telescope} may be
  able to detect the progenitors of GN-z11 analogues out to $z{\sim}$13--14,
  pushing the frontiers of galaxy-formation observations to the early phases of
  cosmic reionization and providing a valuable glimpse of the first galaxies to
  reionize the Universe on large scales.
\end{abstract}




\section{Introduction}

Using \textit{Hubble Space Telescope} (\textit{HST}) Wide Field Camera 3 (WFC3) grism
spectroscopy, \citet{Oesch2016} recently identified the most distant galaxy
known to date (GN-z11).  The spectrum, combined with photometric data
from the Cosmic Assembly Near-infrared Deep Extragalactic Legacy Survey
(CANDELS) survey, placed the system at $z$=11.09$^{+0.08}_{-0.12}$, with lower
redshift spectral energy distribution (SED) solutions excluded with high
confidence.

At $M_\textrm{UV}{=}{-}22.1\pm{}0.2$, GN-z11 is approximately three times more
luminous than $L^*$ at $z$=7--8, and is therefore extremely bright for such a
high-$z$ object.  Extrapolations of the $z$=4--8 ultraviolet luminosity
function (UV LF) suggest that such objects should be rare with fewer than
$\sim$0.06 such systems expected in the total volume surveyed
\citep[${\sim}$1.2$\times$10$^6$\,Mpc$^3$; e.g.][]{Bouwens2015,
Finkelstein2015}.  LF evolution models based on hierarchical assembly
\citep[e.g.][]{Trenti2010, Mason2015}, the extrapolation of abundance matching
results to higher luminosities/redshifts \citep[e.g.][]{Trac2015, Mashian2016},
and cosmological hydrodynamic simulations such as \textsc{BlueTides}
\citep{Waters2016, Waters2016b}, all similarly point to GN-z11 being an
exceptional and rare system.  This begs the question of what the formation
mechanisms might be for such a galaxy, as well as what its descendants are at
lower redshifts?

In this work we use the semi-analytic galaxy-formation model \Meraxes{}
\citep{Mutch2015}, developed for the Dark-ages Reionization And
Galaxy-formation Observables from Numerical Simulations (DRAGONS) programme
\citep{Angel2016,Geil2015,Liu2015}, in order to investigate whether
galaxy-formation models of this type predict the existence of such extreme
systems as GN-z11 and, if so, what their properties, origins and potential
fates are.  The DRAGONS parent \textit{N}-body simulation \citep[\Tiamat{};
][]{Poole2016} possesses a comparable volume to that of the discovery
observations, yet \Meraxes{} predicts two $z$=11.1 galaxies of comparable UV
magnitude to GN-z11, suggesting that such luminous galaxies may be a more
common outcome of early galaxy formation than previously thought.  

This paper is laid out as follows.  In Section~\ref{sec:dragons}, we give a
brief overview of the DRAGONS framework, including both the \Tiamat{}
\textit{N}-body simulation and \Meraxes{} semi-analytic model.  In
Section~\ref{sec:analogues}, we discuss the presence and properties of luminous
GN-z11 analogues in our model galaxy population before going on to investigate
the origin and fate of these objects from $z$=5--20.  In
Section~\ref{sec:JWST}, we discuss the prospects for detecting the progenitors
of GN-z11 analogues with the forthcoming James Webb Space Telescope
(\textit{JWST}) and provide forecasts for the maximum redshift out to which
these may be observed.  We close with a summary of our findings in
Section~\ref{sec:conclusions}.

The \Tiamat{} simulation was run with a standard, spatially flat $\Lambda$CDM
cosmology, utilizing the latest \citet{Planck-Collaboration2015} parameters
($h$, $\Omega_{\rm{m}}$, $\Omega_{\rm{b}}$, $\Omega_\Lambda$, $\sigma_8$,
$n_{\rm{s}}$) = (0.678, 0.308, 0.0484, 0.692, 0.815, 0.968).  However, for
consistency with \citet{Oesch2016} we present all results with $h$=0.7
throughout.

\section{The DRAGONS framework}
\label{sec:dragons}

The DRAGONS programme was designed specifically to study the Epoch of
Reionization (EoR) and the growth of the first galaxies.  It consists of two
main parts: the \Tiamat{} suite of high resolution \textit{N}-body simulations,
and the \Meraxes{} semi-analytic galaxy formation model.

The main \Tiamat{} simulation consists of a (96.9\,Mpc)$^3$ volume with enough
mass resolution to resolve dark matter haloes down to approximately three times
the atomic cooling mass threshold at $z$=5.  It also possesses a high temporal
resolution of $\sim$11\,Myr per output snapshot, allowing us to capture the
stochastic nature of star formation and quenching during the EoR.  Only trees
constructed from the main \Tiamat{} \textit{N}-body simulation are utilized in
this work.  For more details, including information on the other simulations in
the \Tiamat{} suite, please see \citet{Poole2016} and \citet{Angel2016}.

The \Meraxes{} semi-analytic model runs atop the merger trees extracted from
\Tiamat{} and includes the dominant physical process thought to shape the
evolution of high-redshift galaxies, including cooling, star formation, metal
enrichment, satellite infall and merger events.  The model also features a
novel delayed supernova feedback and enrichment scheme and self-consistently
incorporates the semi-numerical reionization scheme, \textsc{21CMFAST}
\citep{Mesinger2007, Mesinger2011}, in order to model reionization and its
effects on galaxy evolution both temporally and spatially \citep{Geil2015}.  A
detailed description of \Meraxes{} and its design can be found in
\citet{Mutch2015}.

The fiducial \Meraxes{} model calibration (used in this work) was constructed
to reproduce the observed evolution of the galaxy stellar mass function from
$z$=5--7 \citep{Gonzalez2011, Duncan2014, Grazian2014, Song2015} and to provide
a reionization history consistent with the latest Planck optical depth
measurements \citep{Planck-Collaboration2015}.  The model has also been shown
to accurately reproduce the evolution of the observed UV LF from $z$=5--10
\citep{Liu2015}.

\section{GN-z11 analogues in DRAGONS}
\label{sec:analogues}

\begin{figure}
  \centering
  \includegraphics[width=0.99\linewidth]{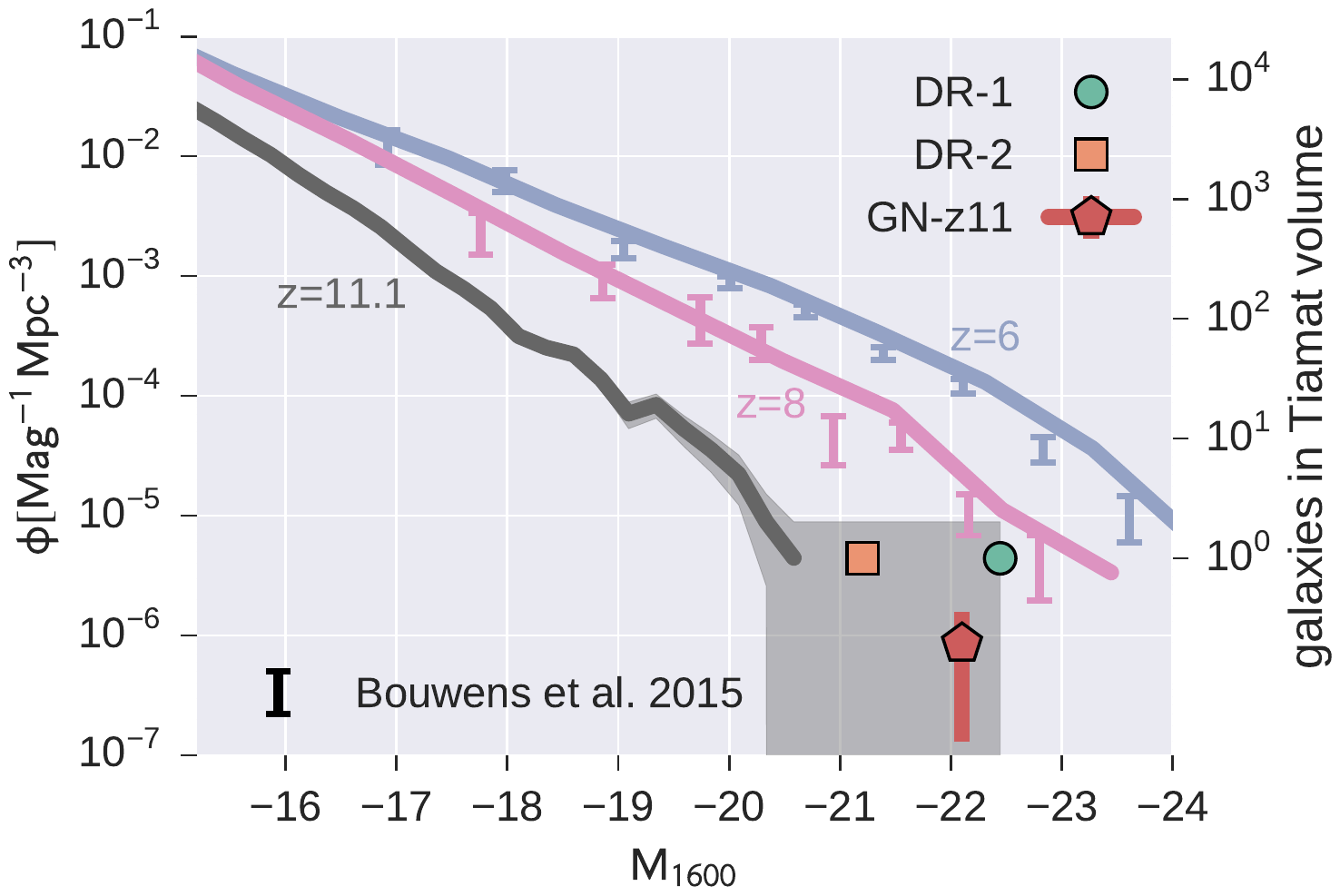}
  \caption{%
    The bright end of the $z$=11.1 UV LF predicted by \Meraxes{} (grey line)
    along with associated Poisson uncertainties (shaded region).  The
    luminosities of our two analogue galaxies, DR-1 and DR-2, are shown as a
    green circle and orange square, respectively.  No dust extinction has been
    applied to the model results.  The red data point indicates observational
    estimated value and 1-$\sigma$ uncertainty derived from the GN-z11
    detection.  Also shown are the $z$=6 and $z$=8 LFs predicted by the model,
    along with observational results from \citet{Bouwens2015} which have been
    dust corrected using the inverse of the methodology outlined in
    \citet{Liu2015}.  The agreement between the model and observations
    demonstrates the success of our model at reproducing the normalization and
    evolution of high-$z$ UV LFs.
  }\label{fig:lf}
\end{figure}

In Fig.~\ref{fig:lf} we present the bright end of the $z$=11.1 UV LF predicted
by \Meraxes{} (grey line and with shaded Poisson uncertainties).  The effects
of dust extinction are not included.  The orange square and green circle
indicate the magnitudes of the two most UV-luminous galaxies in our model
whilst the red data point indicates the observationally estimated $\phi$ and
1$\sigma$ uncertainty derived from the GN-z11 detection.  As can be seen, these
two most luminous \Meraxes{} systems are comparable in luminosity to GN-z11.

\begin{table*}
\begin{minipage}{\textwidth}
  \centering
  \caption{%
    The properties of the GN-z11 analogues, DR-1 and DR-2, at both $z$=11.1 and
    5.  For comparison, the observationally determined values for GN-z11 itself
    are also shown \citep{Oesch2016}.
  }\label{tab:properties}
  \begin{tabular}{lcccccc}
    \hline
    & \multicolumn{3}{c}{$z$=11.1} && \multicolumn{2}{c}{$z$=5} \\
    \cline{2-4} \cline{6-7}
    & GN-z11 & DR-1 & DR-2 && DR-1 & DR-2 \\
    \hline
    $M_\textrm{UV}$ & $-22.1 \pm{} 0.2$ & $-22.4$ & $-21.2$ && $-24.3$ & $-23.2$\\
    $\log_{10} (M_*/$\Msol$)$ & $9.0 \pm{} 0.4$ & $9.3$ & $9.0$ && $11.0$ & $10.6$\\
    $\log_{10} (M_\textrm{vir}/$\Msol$)$ & -- & $11.1$ & $10.8$ && $12.2$ & $11.5$\\
    SFR\,$[$\Msol$/\textrm{yr}^{-1}]$ & $24.0 \pm{} 10$ & $66.4$ & $19.0$ && $237.6$ & $94.5$\\
    $r_\textrm{half-light}$\,[kpc] & $0.6 \pm{} 0.2$ & $0.3$ & $0.2$ && $0.9$ & $0.4$\\
    \hline
  \end{tabular}
\end{minipage}
\end{table*}

For reference, we also show in Fig.~\ref{fig:lf} the $z$=6 and 8 UV LFs
predicted by \Meraxes{} as well as the observational results of
\citet{Bouwens2015}.  The observations have been corrected for dust extinction
using the inverse of the methodology described in \citet{Liu2015}.  The
agreement between the observational data and our model results indicates that
\Meraxes{} correctly predicts both the normalization and evolution of the
bright end of the UV LF at lower redshifts, giving us confidence in the
$z$=11.1 result.

Due to the limited volume of the parent \Tiamat{} \textit{N}-body simulation,
the bright end of the LF is poorly constrained for number densities
${\lesssim}10^{-5}$\,Mag$^{-1}$Mpc$^{-3}$.  This precludes us from making any
quantitative statements regarding the probability of detecting a galaxy as
bright as GN-z11 other than to simply state that we find at least one such
system in our ${\sim}$1$\times$10$^6$\,Mpc$^3$ volume.  However, we note that
this is approximately equal to the volume surveyed during the detection of
GN-z11 (${\sim}$1.2$\times$10$^6$\,Mpc$^3$) and therefore the presence of any
galaxies as bright as this observed object in our model is significant.  Given
the success of \Meraxes{} in reproducing the evolution of the galaxy stellar
mass functions \citep{Mutch2015} and observed UV LFs \citep{Liu2015} at
$z$=5--8, the presence of any systems as luminous as GN-z11 in our model is
remarkable, opening up the possibility that such galaxies may not be as rare as
extrapolations of the observed $z{\sim}$4--8 LFs to higher redshifts may
suggest \citep[see][and references therein]{Oesch2016}.

Having established that galaxies as UV-luminous as GN-z11 exist in the output
of DRAGONS, we selected the two brightest UV magnitude galaxies at redshift
$z$=11.1 for detailed study.  The properties of these
objects,\footnote{\label{foot:r_half-light}The model galaxy half-light radii are
  calculated from the disc scale radius, $r_\textrm{s}$, using
  $r_\textrm{half-light}{=}1.68 r_\textrm{s}$ for an axisymmetric exponential
  disc profile, where $r_\textrm{s}$ is derived from the spin of the host dark
matter subhalo under the assumption of specific angular momentum conservation
\citep{Fall1980, Mo1998, Mutch2015}.\\ \\ } hence forth referred to as DR-1
and DR-2, show good agreement with the best observational measurements of
GN-z11 \citep{Oesch2016}, as can be seen in Table~\ref{tab:properties}.  This
suggests that we can use these model galaxies as analogues with which to
investigate the potential formation, evolution and fate of GN-z11.

\begin{figure}
  \centering
  \includegraphics[width=0.99\linewidth]{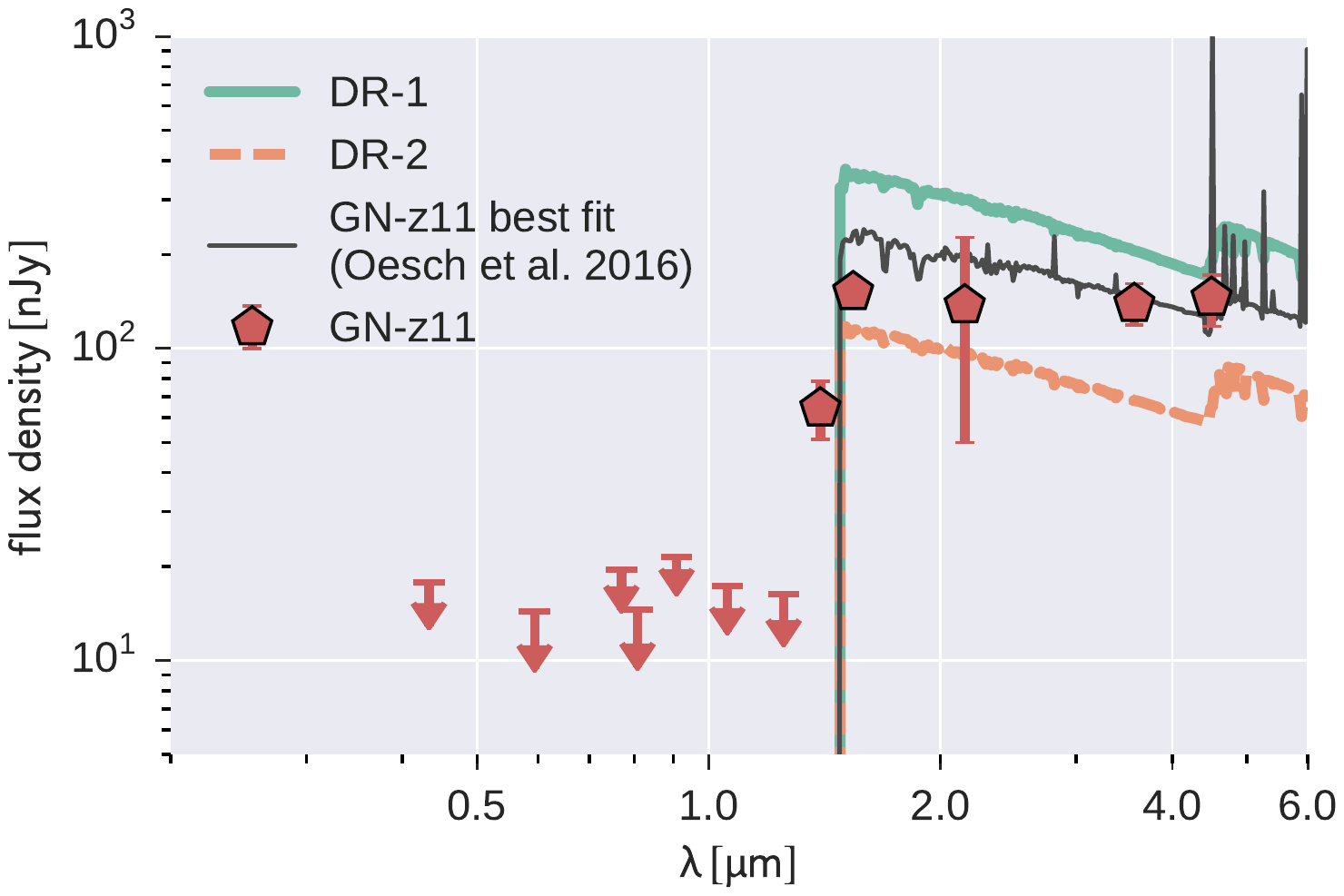}
  \caption{%
    The observed frame SED of DR-1 (green) and DR-2 (orange) in units of flux
    density.  Lyman-$\alpha$ absorption has been included, however, no dust
    extinction model has been applied.  Red points show the observed GN-z11
    \textit{HST} photometric measurements and upper limits.  The black thin
    line indicates the best-fitting SED for GN-z11 \citep{Oesch2016}.  Both
    model analogue galaxies have a spectral shape in good agreement with the
    GN-z11 measurements. The normalization of both model spectra further show
    reasonable agreement with GN-z11 photometric measurements due to their
    comparable UV luminosities (see Table~\ref{tab:properties}).
  }\label{fig:sed}
\end{figure}

As well as those properties listed in Table~\ref{tab:properties}, DR-1 and DR-2
also possess similar SEDs to GN-z11.  In Fig.~\ref{fig:sed} we plot the
observed frame SEDs of DR-1 (green) and DR-2 (orange) in terms of their flux
density between 0.2 and 6\,$\upmu$m.  The red data points show the GN-z11
\textit{HST} photometric measurements and upper limits, whilst the black thin
line indicates the best-fitting SED presented in \citet{Oesch2016}.  Lyman-$\alpha$
absorption has been included in the model spectra and is manifested as the
sharp drop in flux at $\lambda{\lesssim}1.1\,{\mu}$m; however, no dust
extinction model has been applied.  For more details on the methodology used to
construct the model SEDs, see \citet{Liu2015}.  Both analogue galaxies possess
spectra and UV slopes ($\beta$) in close agreement with the GN-z11
observations, supporting the claim that this observed system is relatively dust
free \citep{Oesch2016}.  The normalization of both model spectra further show
reasonable agreement with GN-z11 photometric measurements due to their
comparable UV luminosities (cf. Table~\ref{tab:properties}).

\begin{figure}
  \centering
  \includegraphics[width=0.99\linewidth]{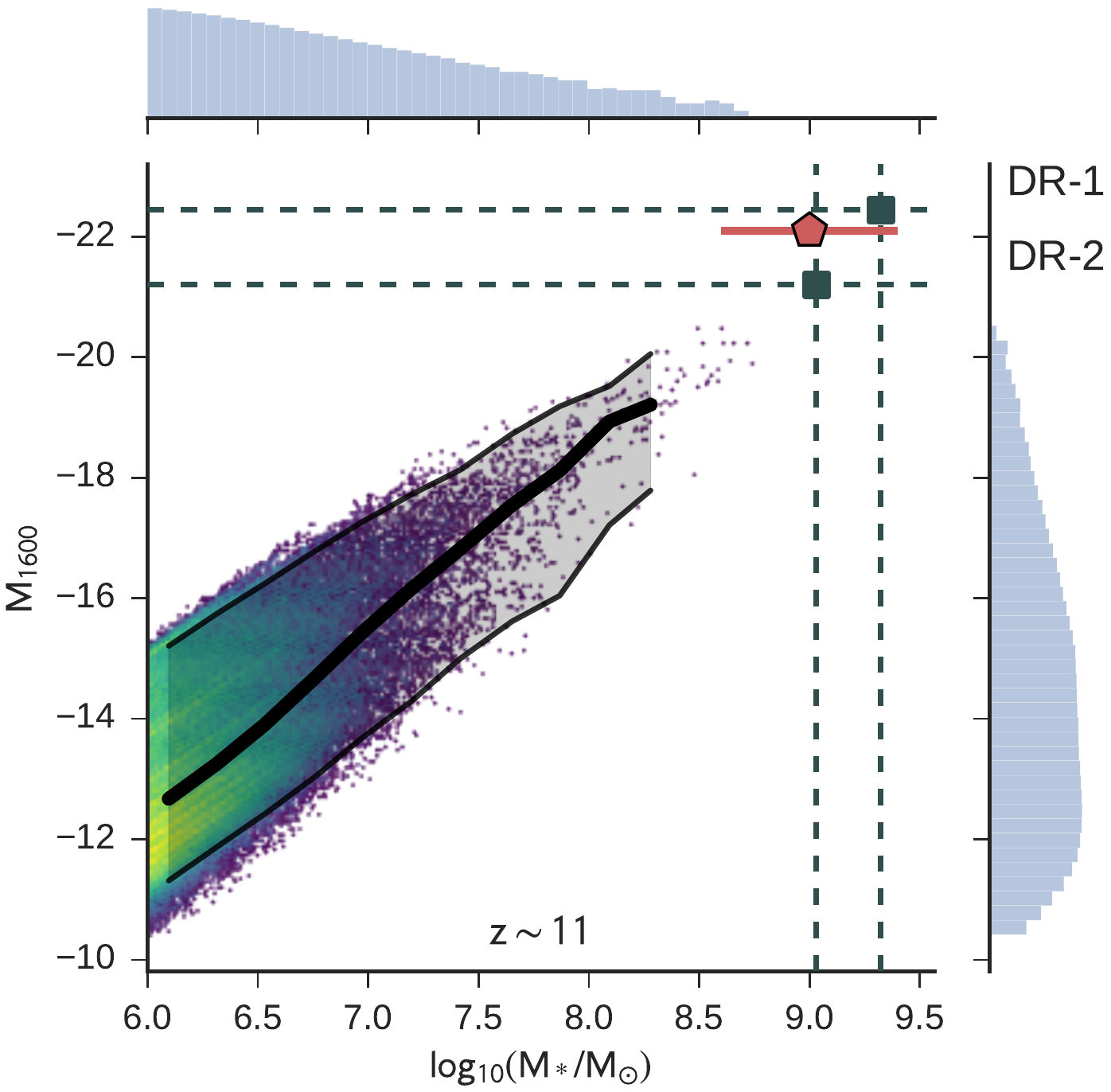}
  \caption{%
    The UV magnitude versus stellar mass distribution of \Meraxes{}
    galaxies at \zelv{}. The thick black line and shaded region indicate the
    median and 95\ pc confidence intervals of the distribution as a function of
    stellar mass.  The top (right) panels indicate the marginalized
    log-probability distribution of stellar mass (UV magnitude) values.  The
    red point with error bars indicates the position of GN-z11 in this plane,
    whilst the grey squares show the model analogue galaxies, DR-1 and DR-2.
    Both analogue galaxies (as well as GN-z11) are rare outliers from the main
    distribution, but are approximately consistent with an extrapolation of the
    median relation from lower masses.
  }\label{fig:mag_vs_stars}
\end{figure}

DR-1 and DR-2 are the two most massive galaxies in the simulation at the redshift
at which they were selected ($z$=11.1) and are hosted by the two most massive
subhaloes.  They are also rare outliers from the majority of the model galaxy
population in terms of their stellar masses, star formation rates and UV
luminosities.  However, they remain broadly consistent with the mean trends
displayed by galaxies at lower luminosities/masses, suggesting that their
history is not particularly special or unique.  As an example, in
Fig.~\ref{fig:mag_vs_stars} we present the distribution of all \Meraxes{}
$z$=11.1 galaxies in the UV magnitude versus stellar mass plane.  The positions of
DR-1 and DR-2 are shown as grey squares, whilst GN-z11 is indicated by the red
point with error bars.  Although these three objects lie out with the bulk of
the main distribution, they are roughly consistent with the median
$M_{1600}$--$M_*$ trend extrapolated from lower masses.

\section{The origin and fate of GN-z11}
\label{sec:origin-and-fate}

The detection of such a massive and luminous galaxy at $z{\sim}$11 raises a
number of interesting questions.  How do such massive systems form so rapidly?
Is their extreme brightness merely a transient feature of their evolution
brought on by a merger or other significant event?  If not, then how can the
level of star formation required to produce new, UV luminous stars be
maintained?  In this section we address these questions using our model
analogue galaxies and further explore what the eventual fate of such objects
is at lower redshifts.

\subsection{Formation history}
\label{sec:formation_history}

\begin{figure}
  \centering
  \includegraphics[width=0.99\linewidth]{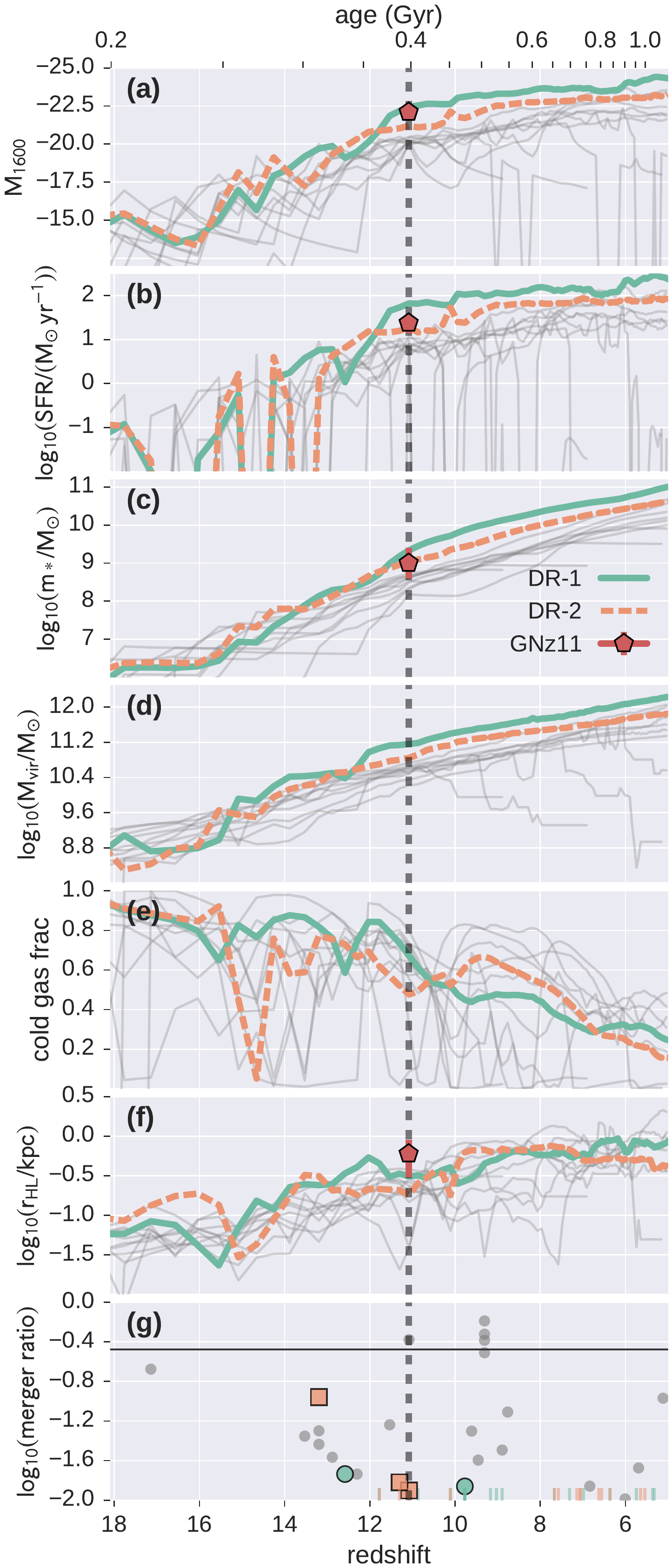}
  \caption{%
    Property histories for the model analogue galaxies, DR-1 (green) and DR-2
    (orange): \textit{(a)} rest frame intrinsic absolute UV magnitude
    (1600\,\AA); \textit{(b)} star formation rate; \textit{(c)} stellar mass;
    \textit{(d)} subhalo mass; \textit{(e)} cold gas fraction; \textit{(f)}
    half light radius assuming a pure-exponential disc; \textit{(g)} merger
    baryonic mass ratios for all galaxy merger events with ratios $>$0.01
    (mergers of lower mass ratio are indicated by tick marks on the $x$-axis).
    The red dots with error bars show the best estimates of the properties of
    GN-z11 \citep{Oesch2016}, whilst the vertical dashed line indicates $z$=11.
    For comparison, thin grey lines and dots show the histories of the ten
    next-most-luminous galaxies selected at the same redshift as our analogues
    ($z{\sim}11$).
  }\label{fig:histories}
\end{figure}

Fig.~\ref{fig:histories} shows the full formation histories of DR-1 (green)
and DR-2 (orange) for a number of properties from $z$=18 to 5.  The corresponding
measured values for GN-z11 are shown as red points with error bars where
available.  As with the UV luminosity and stellar masses discussed above, we
see close agreement between our model analogues and GN-z11 in terms of star
formations rates (panel \textit{b}) and disc sizes (panel \textit{f}).

The UV luminosities of both DR-1 and DR-2 continually, but extremely rapidly,
increase from high redshift (panel \textit{a}).  The formation (or half-mass)
redshift of each $z$=11.1 analogue's dark matter subhalo is $z{\sim}$11.4 and
11.9, respectively.  For comparison, theoretical expectations using extended
Press--Schechter theory \citep{Bond1991} are for these two haloes to have formed
far earlier, at $z{\sim}$12.1, and hence to have grown at a much slower pace
\citep{Tacchella2013, Trenti2015}.  In our simulation, the host subhalo of DR-1
grows by a factor of approximately 5 in just 65\,Myr (panel \textit{c})
immediately preceding $z$=11.1, resulting in a growth in stellar mass of a
factor of 9 during the same period.  A visual inspection of the evolution of
the local environment of DR-1 indicates that this period of rapid growth
coincides with infall into a filamentary like structure.  During this time, the
host subhalo is subjected to multiple minor mergers (of haloes with no stellar
mass), as well as significant smooth accretion below the resolution limit of
the simulation.

It is also immediately apparent from Fig.~\ref{fig:histories}, however, that
the extreme UV brightness of DR-1 and DR-2 are not transient features of their
evolution; the star formation rates and UV luminosities of both analogue
systems remain high or increasing throughout their evolution.  For comparison,
thin grey lines show the evolution of the 10 next-most-luminous galaxies
selected at the same redshift ($z$=11.1).  These luminous counterpart galaxies
all possess similar growth histories, with high star formation rates (panel
\textit{b}) driven by continual growth of their host haloes (panel \textit{d})
and associated accretion of fresh gas.  However, their star formation is less
sustained than that of DR-1 and 2, with more numerous periods of inactivity.

\begin{figure*}
\begin{minipage}{\textwidth}
  \centering
  \includegraphics[width=0.99\columnwidth]{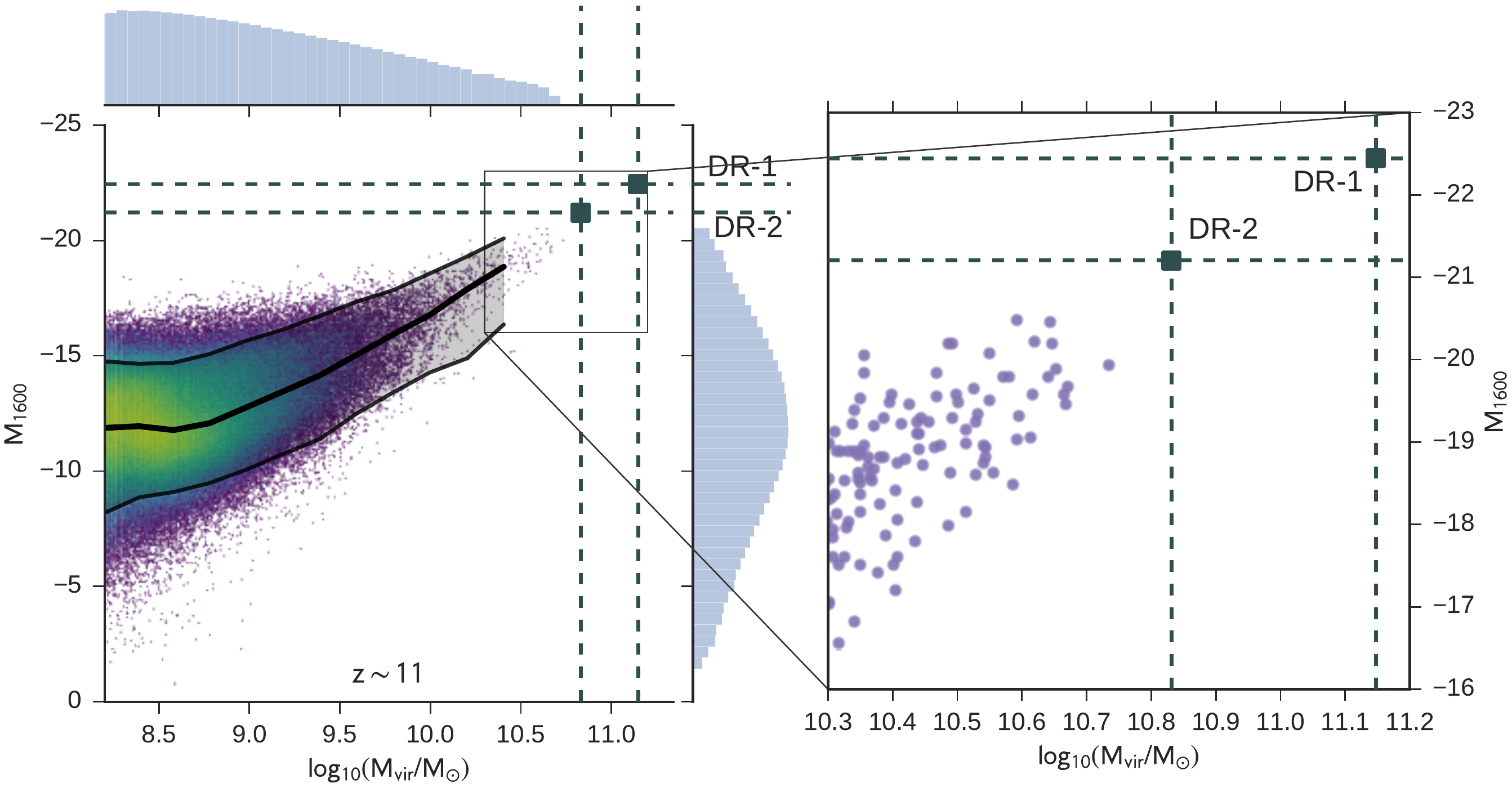}\\
  \includegraphics[width=0.99\columnwidth]{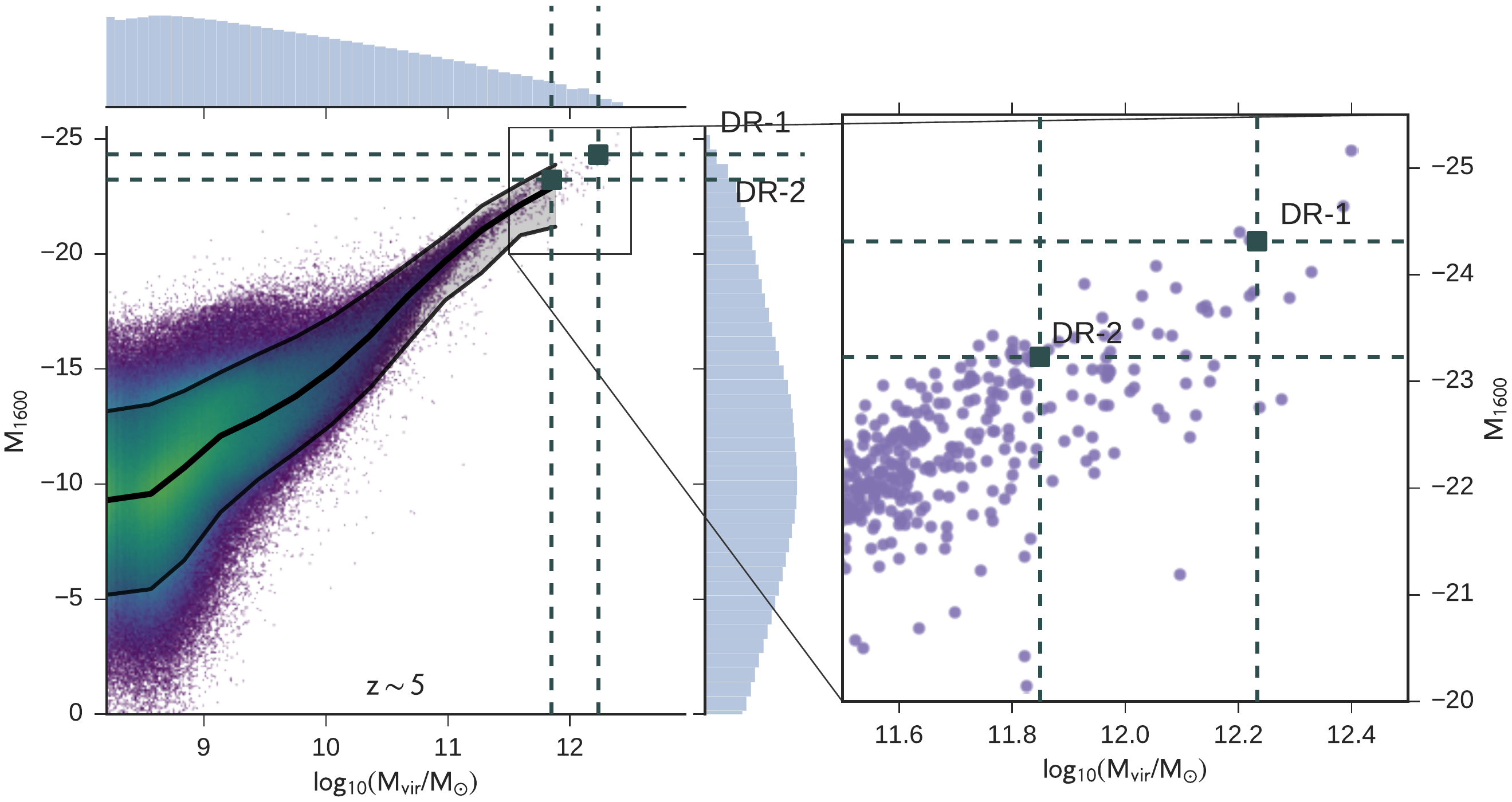}
  \caption{%
    The distribution of model galaxy UV magnitudes (M$_{1600}$) as a function
    of subhalo mass M$_\textrm{vir}$ at $z{\sim}$11 (top row).  The right-hand
    panel shows a zoom in scatter plot of the highest M$_\textrm{vir}$
    (/M$_{1600}$) region.  Again, both GN-z11 analogue galaxies are outliers of
    the main population.  However, by $z$=5 (bottom row), this is no longer the
    case.  Whilst these galaxies remain some of the most UV luminous and
    massive, they have regressed towards the mean trend and there now exists a
    small number of other, more extreme systems.
  }\label{fig:mag_vs_mvir}
\end{minipage}
\end{figure*}

So how can DR-1 and 2 achieve and sustain such a rapid stellar mass growth?
High star formation rates should result in large amounts of supernova energy
being deposited into the interstellar medium, heating and ejecting cold gas and
thus curtailing star formation until this material is replenished.  Indeed,
this is the dominant mechanism by which star formation is regulated in
\Meraxes{} \citep{Mutch2015}.  In panel \textit{(e)} of
Fig.~\ref{fig:histories} we show the cold gas fraction, $M_\textrm{cold}/
(m_\textrm{cold}{+}M_*)$, of our ten luminous comparison galaxies in grey.
Many show the expected large drops in cold gas fraction caused by star
formation events which deplete gas, both by converting it to stars and ejecting
it from the galaxy through supernova feedback.  Occasionally these low gas
fractions remain for tens of millions of years, but more often gas reserves are
quickly replenished by accretion from the IGM\@.  By comparison, our GN-z11
analogues, DR-1 and DR-2, show typically fewer major depletion events.  This is
due to the large amounts of cold gas available to these galaxies which results
in any one star formation episode removing only a small fraction of the
available material.

In panel \textit{(g)} we show the baryonic merger ratios
($m_\textrm{baryon}^\textrm{satellite}/m_\textrm{baryon}^\textrm{central}$) of
all merger events with mass ratios greater than 0.01 for both DR-1 and DR-2, as
well as our ten comparison objects.  The horizontal grey line indicates a ratio
of 1/3, above which we deem the merger to be a major event.  The ticks on the
lower $x$-axis indicate merger events with ratios below the range of the plot.
Interestingly, we see that neither DR-1 nor DR-2 has experienced a merger with a
baryonic mass ratio greater than $\sim$0.1.  We further find that none of the
luminous comparison galaxies has experienced a major merger prior to \zelv{}.
Although these results refer to galaxy merger events (as opposed to mergers
between haloes which may potentially be devoid of stars), we confirm that
$>$90\% of the $z{<}$11.1 halo mass growth of all plotted objects is driven by
smooth accretion of systems below the halo mass resolution limit of our input
\textit{N}-body simulation (${\sim}$1.4$\times 10^8$\,$h^{-1}$M$_{\sun}$).

Such quiescent merger histories are an important aspect of the successful
growth of GN-z11 analogues in \Meraxes{}.  Within our model, and as suggested
by hydrodynamical simulations \citep[e.g.][]{Cox2008,Powell2013,Kannan2015},
major merger events induce shocks and instabilities which in turn result in
efficient star formation events.  Such events consume large fractions of the
cold gas of both progenitors with the resulting supernova feedback ejecting
even more material from the system.  Although both DR-1 and 2 have plentiful
and continuous replenishment of cold gas from the IGM, it is likely that
numerous major merger events would have curtailed the eventual \zelv{} stellar
mass and would certainly have introduced more variability into the star
formation histories of these objects, reducing their duty cycles and limiting
their probability of detection.

\subsection{Eventual fate}
\label{sec:fate}

As well as using the DRAGONS framework to explore the origins of GN-z11
analogues, we can also exploit the realistic galaxy populations it provides to
explore the eventual fate of these rare objects.  From
Fig.~\ref{fig:histories}, we can see that the haloes which DR-1 and 2 occupy
continue to grow steadily to $z$=5.  This steady growth brings with it fresh
gas, allowing for a slowly increasing star formation rate and UV luminosity.
As such, these galaxies remain the most luminous and massive of the twelve
objects shown in the figure (selected to be the twelve most UV-luminous objects
at $z$=11.1).  However, when compared to the full galaxy population at $z$=5,
our two GN-z11 analogues are no longer the brightest or most massive in the
simulation volume.  Such a regression of extreme objects towards the mean at
later times is an expected feature of structure formation \citep{Trenti2008}. 

In Fig.~\ref{fig:mag_vs_mvir} we show the absolute UV magnitude versus subhalo
mass distribution for the full galaxy population at $z$=11.1 (top) and $z$=5
(bottom).  The panels on the right-hand side show a zoom in scatter plot of the
highest $M_\textrm{vir}$ (/$M_{1600}$) regions.  As with
Fig.~\ref{fig:mag_vs_stars} above, the thick black line indicates the median
of the distribution, with the shaded region showing the 95\,pc confidence
intervals.  Grey squares indicate the positions of DR-1 and DR-2.  At \zelv{} both
galaxies lie in the most massive subhaloes and are clear outliers from the bulk
of the galaxy population.  At $z$=5, this is no longer the case.  Not only have
both systems rejoined the tail of the main distribution, but there are several
systems which are both more massive and luminous.  Our model therefore predicts
that although GN-z11 may be a highly biased, rare galaxy in the redshift
\zelv{} Universe, it is likely the progenitor of more common massive galaxies
in the post-reionization epoch.  For reference, we present the full $z$=5
properties of DR-1 and DR-2\ in Table~\ref{tab:properties}.

\section{Detectability and prospects for \textit{JWST}}
\label{sec:JWST}

\begin{figure}
  \centering
  \includegraphics[width=0.99\linewidth]{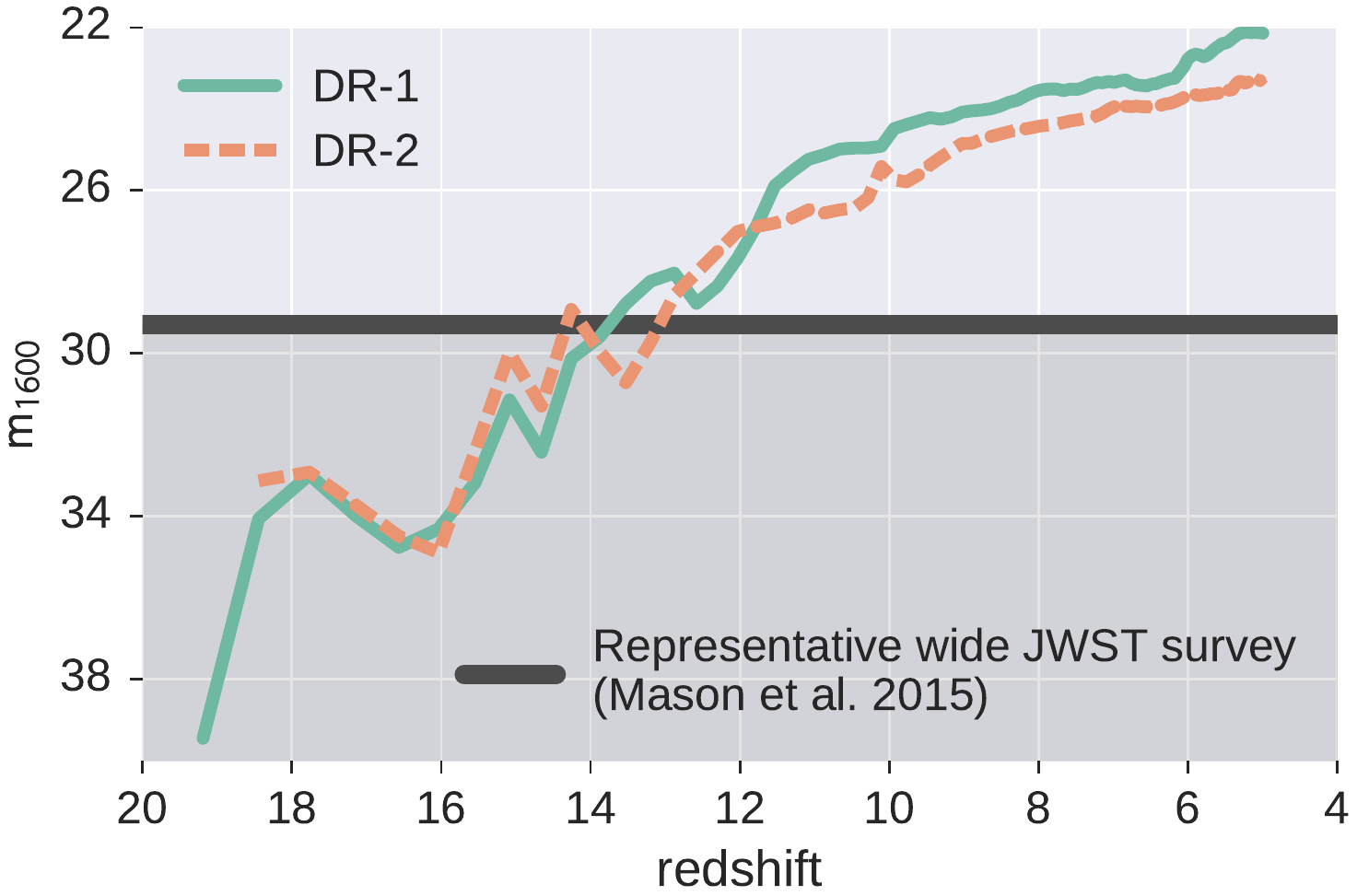}
  \caption{%
    The rest-frame intrinsic UV apparent magnitude history of DR-1 (green) and
    DR-2 (orange) along with the 8$\sigma$ detection threshold (grey horizontal
    line) corresponding to the example future wide-field \textit{JWST}/NIRCAM
    survey presented in \citet{Mason2015}.  This survey design corresponds to
    an 800 hour total exposure time split amongst multiple pointings and bands,
    covering a total survey area of 4000\,arcmin$^2$.  See
    Section~\ref{sec:JWST} for more details.  Our model predicts that future
    \textit{JWST} surveys such as this may be able to detect the progenitors of
    GN-z11 analogues out to $z{\sim}$13--14, corresponding to the early
    stages of reionization.
  }\label{fig:mapp_histories}
\end{figure}

GN-z11 represents the highest redshift galaxy found to date, and
pushes the boundaries of what is technically achievable using \textit{HST}.
\textit{JWST} will possess the capability to go approximately two magnitudes
deeper in the rest frame UV at high redshift, opening up the possibility of
detecting more GN-z11 type systems at $z{\sim}$11 \citep{Waters2016}, as well
as their progenitors at higher redshifts.

In Fig.~\ref{fig:mapp_histories} we present the rest-frame apparent UV
magnitude histories of DR-1 and DR-2 from $z$=5--20 (green and orange lines,
respectively).  No dust extinction has been included, in agreement with the
observed blue UV continuum slope of GN-z11 \citep{Oesch2016}.  The grey
horizontal line indicates the 8$\sigma$ detection limit for the example
wide-field \textit{JWST}/NIRCAM survey presented in \citet{Mason2015}.  This
survey design corresponds to a total of 800 hours of exposure time divided up
amongst 400 pointings in 5 bands, and covering a total area of $\sim$4000
arcmin$^2$.  Further details can be found in \citet{Mason2015}.

The total comoving volume of this example survey is (178.4\,Mpc)$^3$ at
$z$=11.1.  This is approximately 6 times the volume of our \Tiamat{}
simulation, and $\sim$90\% of that of the much larger \textsc{BlueTides}
simulation \citep{Feng2015}, within which it has been recently shown that
approximately 30 systems of similar or brighter luminosity than GN-z11 are
found \citep{Waters2016}.  Such a survey should therefore provide both a volume
and sensitivity to easily detect multiple GN-z11 analogue systems at
$z{\sim}$11.  Our model further suggests that the progenitors of such systems
may remain luminous enough to be detected out to $z{\sim}$13--14, corresponding
to the early stages of Reionization when the Universe was less than $\sim$10\%
ionized \citep[e.g.][Greig \& Mesinger in prep.]{Kuhlen2012,Mutch2015}.

\section{Conclusions}
\label{sec:conclusions}

In this work, we identify two GN-z11 analogue systems from the results of
the semi-analytic model, \Meraxes{}, created as part of the DRAGONS programme.
The presence of these two analogues in a simulated volume approximately equal to
that of the observational survey volume supports claims by other authors
\citep{Waters2016} that galaxies as luminous as GN-z11 are more common than
extrapolations of $z{\sim}$4--8 UV LFs would suggest \citep{Oesch2016}.

Using the detailed properties and full formation histories afforded to us by
\Meraxes{}, we have investigated the formation, evolution and fate of our two
model analogue systems (DR-1 and 2).  Our results can be summarized as follows.

\begin{itemize}
  \item Both DR-1 and 2 possess stellar masses, UV luminosities, star formation
    rates, sizes and SEDs, which are in close agreement with the measured values
    of GN-z11 (Section~\ref{sec:analogues}).
  \item These analogues are the most UV-luminous and massive systems in our
    simulated volume at $z$=11 and are rare outliers from the bulk of the
    galaxy population.  However, despite their extreme nature, their properties
    are broadly consistent with extrapolations of the mean trends at lower
    masses (Section~\ref{sec:analogues}).
  \item The extreme UV brightness of our analogues at $z=11$ is not a transient
    feature of their histories.  Instead both their luminosities and stellar
    masses increase relatively smoothly, but extremely rapidly, from higher
    redshifts and continue to increase for the remainder of their evolution to
    $z$=5 (Section~\ref{sec:formation_history}).
  \item The sustained and increasing SFRs of DR-1 and DR-2 are due to rapid growth
    of their host dark matter haloes, driven predominantly by smooth accretion
    of objects below the mass resolution limit of our simulation and bringing
    with it an ample supply of fresh cold gas to fuel star formation.  Another
    important feature of these growth histories is a notable lack of major
    mergers which would have caused a significant star burst, ejecting a large
    amount of cold gas and temporarily stalling star formation
    (Section~\ref{sec:formation_history}).
  \item Despite being the most extreme and rare outliers of the full galaxy
    population at $z{\sim}$11, by $z$=5 neither analogue is the most massive nor
    UV-luminous system in the simulation (Section~\ref{sec:fate}).
  \item Future, wide-field surveys with \textit{JWST} will likely be able to
    identify the progenitors of GN-z11 type galaxies out to
    $z{\sim}$13--14 (Section~\ref{sec:JWST}).
\end{itemize}

The potential ability of \textit{JWST} to detect GN-z11 progenitors all the way
out to $z{\sim}$14 will push the frontiers of galaxy-formation observations to
the early phases of cosmic reionization and provide a valuable glimpse of the
first galaxies to reionize the Universe on large scales.

\section*{Acknowledgements}

This research was supported by the Victorian Life Sciences Computation
Initiative (VLSCI), grant ref. UOM0005, on its Peak Computing Facility hosted
at the University of Melbourne, an initiative of the Victorian Government,
Australia. Part of this work was performed on the gSTAR national facility at
Swinburne University of Technology.\ gSTAR is funded by Swinburne and the
Australian Governments Education Investment Fund. AM acknowledges support from
the European Research Council (ERC) under the European Union's Horizon
2020 research and innovation programme (grant agreement No 638809 AIDA). PO and
GDI acknowledge the support of NASA grant HST-GO-1387 awarded by the Space
Telescope Science Institute, which is operated by the Association of
Universities for Research in Astronomy, Inc., under NASA contract NAS 5\-26555.
The DRAGONS research programme is funded by the Australian Research Council
through the ARC Laureate Fellowship FL110100072 awarded to JSBW.\




\bibliographystyle{mnras}
\bibliography{gnz11} 

\begin{thebibliography}{}
\makeatletter
\relax
\def\mn@urlcharsother{\let\do\@makeother \do\$\do\&\do\#\do\^\do\_\do\%\do\~}
\def\mn@doi{\begingroup\mn@urlcharsother \@ifnextchar [ {\mn@doi@}
  {\mn@doi@[]}}
\def\mn@doi@[#1]#2{\def\@tempa{#1}\ifx\@tempa\@empty \href
  {http://dx.doi.org/#2} {doi:#2}\else \href {http://dx.doi.org/#2} {#1}\fi
  \endgroup}
\def\mn@eprint#1#2{\mn@eprint@#1:#2::\@nil}
\def\mn@eprint@arXiv#1{\href {http://arxiv.org/abs/#1} {{\tt arXiv:#1}}}
\def\mn@eprint@dblp#1{\href {http://dblp.uni-trier.de/rec/bibtex/#1.xml}
  {dblp:#1}}
\def\mn@eprint@#1:#2:#3:#4\@nil{\def\@tempa {#1}\def\@tempb {#2}\def\@tempc
  {#3}\ifx \@tempc \@empty \let \@tempc \@tempb \let \@tempb \@tempa \fi \ifx
  \@tempb \@empty \def\@tempb {arXiv}\fi \@ifundefined
  {mn@eprint@\@tempb}{\@tempb:\@tempc}{\expandafter \expandafter \csname
  mn@eprint@\@tempb\endcsname \expandafter{\@tempc}}}

\bibitem[\protect\citeauthoryear{{Angel}, {Poole}, {Ludlow}, {Duffy}, {Geil},
  {Mutch}, {Mesinger}  \& {Wyithe}}{{Angel} et~al.}{2016}]{Angel2016}
{Angel} P.~W.,  {Poole} G.~B.,  {Ludlow} A.~D.,  {Duffy} A.~R.,  {Geil} P.~M.,
  {Mutch} S.~J.,  {Mesinger} A.,   {Wyithe} J.~S.~B.,  2016, \mn@doi [\mnras]
  {10.1093/mnras/stw737}, \href
  {http://adsabs.harvard.edu/abs/2016MNRAS.459.2106A} {459, 2106}

\bibitem[\protect\citeauthoryear{{Bond}, {Cole}, {Efstathiou}  \&
  {Kaiser}}{{Bond} et~al.}{1991}]{Bond1991}
{Bond} J.~R.,  {Cole} S.,  {Efstathiou} G.,   {Kaiser} N.,  1991, \mn@doi
  [\apj] {10.1086/170520}, \href
  {http://adsabs.harvard.edu/abs/1991ApJ...379..440B} {379, 440}

\bibitem[\protect\citeauthoryear{{Bouwens} et~al.,}{{Bouwens}
  et~al.}{2015}]{Bouwens2015}
{Bouwens} R.~J.,  et~al., 2015, \mn@doi [\apj] {10.1088/0004-637X/803/1/34},
  \href {http://adsabs.harvard.edu/abs/2015ApJ...803...34B} {803, 34}

\bibitem[\protect\citeauthoryear{{Cox}, {Jonsson}, {Somerville}, {Primack}  \&
  {Dekel}}{{Cox} et~al.}{2008}]{Cox2008}
{Cox} T.~J.,  {Jonsson} P.,  {Somerville} R.~S.,  {Primack} J.~R.,   {Dekel}
  A.,  2008, \mn@doi [\mnras] {10.1111/j.1365-2966.2007.12730.x}, \href
  {http://adsabs.harvard.edu/abs/2008MNRAS.384..386C} {384, 386}

\bibitem[\protect\citeauthoryear{Duncan et~al.,}{Duncan
  et~al.}{2014}]{Duncan2014}
Duncan K.,  et~al., 2014, MNRAS, 444, 2960

\bibitem[\protect\citeauthoryear{{Fall} \& {Efstathiou}}{{Fall} \&
  {Efstathiou}}{1980}]{Fall1980}
{Fall} S.~M.,  {Efstathiou} G.,  1980, \mn@doi [\mnras]
  {10.1093/mnras/193.2.189}, \href
  {http://adsabs.harvard.edu/abs/1980MNRAS.193..189F} {193, 189}

\bibitem[\protect\citeauthoryear{{Feng}, {Di Matteo}, {Croft}, {Tenneti},
  {Bird}, {Battaglia}  \& {Wilkins}}{{Feng} et~al.}{2015}]{Feng2015}
{Feng} Y.,  {Di Matteo} T.,  {Croft} R.,  {Tenneti} A.,  {Bird} S.,
  {Battaglia} N.,   {Wilkins} S.,  2015, \mn@doi [\apjl]
  {10.1088/2041-8205/808/1/L17}, \href
  {http://adsabs.harvard.edu/abs/2015ApJ...808L..17F} {808, L17}

\bibitem[\protect\citeauthoryear{{Finkelstein} et~al.,}{{Finkelstein}
  et~al.}{2015}]{Finkelstein2015}
{Finkelstein} S.~L.,  et~al., 2015, \mn@doi [\apj]
  {10.1088/0004-637X/810/1/71}, \href
  {http://adsabs.harvard.edu/abs/2015ApJ...810...71F} {810, 71}

\bibitem[\protect\citeauthoryear{{Geil}, {Mutch}, {Poole}, {Angel}, {Duffy},
  {Mesinger}  \& {Wyithe}}{{Geil} et~al.}{2016}]{Geil2015}
{Geil} P.~M.,  {Mutch} S.~J.,  {Poole} G.~B.,  {Angel} P.~W.,  {Duffy} A.~R.,
  {Mesinger} A.,   {Wyithe} J.~S.~B.,  2016, \mn@doi [\mnras]
  {10.1093/mnras/stw1718}, \href
  {http://adsabs.harvard.edu/abs/2016MNRAS.462..804G} {462, 804}

\bibitem[\protect\citeauthoryear{Gonz{\'a}lez, Labbe, Bouwens, Illingworth,
  Franx  \& Kriek}{Gonz{\'a}lez et~al.}{2011}]{Gonzalez2011}
Gonz{\'a}lez V.,  Labbe I.,  Bouwens R.~J.,  Illingworth G.,  Franx M.,   Kriek
  M.,  2011, ApJ Letters, 735, L34

\bibitem[\protect\citeauthoryear{{Grazian} et~al.,}{{Grazian}
  et~al.}{2015}]{Grazian2014}
{Grazian} A.,  et~al., 2015, \mn@doi [\aap] {10.1051/0004-6361/201424750},
  \href {http://adsabs.harvard.edu/abs/2015A%26A...575A..96G} {575, A96}

\bibitem[\protect\citeauthoryear{{Kannan}, {Macci{\`o}}, {Fontanot}, {Moster},
  {Karman}  \& {Somerville}}{{Kannan} et~al.}{2015}]{Kannan2015}
{Kannan} R.,  {Macci{\`o}} A.~V.,  {Fontanot} F.,  {Moster} B.~P.,  {Karman}
  W.,   {Somerville} R.~S.,  2015, \mn@doi [\mnras] {10.1093/mnras/stv1633},
  \href {http://adsabs.harvard.edu/abs/2015MNRAS.452.4347K} {452, 4347}

\bibitem[\protect\citeauthoryear{{Kuhlen} \& {Faucher-Gigu{\`e}re}}{{Kuhlen} \&
  {Faucher-Gigu{\`e}re}}{2012}]{Kuhlen2012}
{Kuhlen} M.,  {Faucher-Gigu{\`e}re} C.-A.,  2012, \mn@doi [\mnras]
  {10.1111/j.1365-2966.2012.20924.x}, \href
  {http://adsabs.harvard.edu/abs/2012MNRAS.423..862K} {423, 862}

\bibitem[\protect\citeauthoryear{{Liu}, {Mutch}, {Angel}, {Duffy}, {Geil},
  {Poole}, {Mesinger}  \& {Wyithe}}{{Liu} et~al.}{2016}]{Liu2015}
{Liu} C.,  {Mutch} S.~J.,  {Angel} P.~W.,  {Duffy} A.~R.,  {Geil} P.~M.,
  {Poole} G.~B.,  {Mesinger} A.,   {Wyithe} J.~S.~B.,  2016, \mn@doi [\mnras]
  {10.1093/mnras/stw1015}, \href
  {http://adsabs.harvard.edu/abs/2016MNRAS.462..235L} {462, 235}

\bibitem[\protect\citeauthoryear{{Mashian}, {Oesch}  \& {Loeb}}{{Mashian}
  et~al.}{2016}]{Mashian2016}
{Mashian} N.,  {Oesch} P.~A.,   {Loeb} A.,  2016, \mn@doi [\mnras]
  {10.1093/mnras/stv2469}, \href
  {http://adsabs.harvard.edu/abs/2016MNRAS.455.2101M} {455, 2101}

\bibitem[\protect\citeauthoryear{{Mason}, {Trenti}  \& {Treu}}{{Mason}
  et~al.}{2015}]{Mason2015}
{Mason} C.~A.,  {Trenti} M.,   {Treu} T.,  2015, \mn@doi [\apj]
  {10.1088/0004-637X/813/1/21}, \href
  {http://adsabs.harvard.edu/abs/2015ApJ...813...21M} {813, 21}

\bibitem[\protect\citeauthoryear{{Mesinger} \& {Furlanetto}}{{Mesinger} \&
  {Furlanetto}}{2007}]{Mesinger2007}
{Mesinger} A.,  {Furlanetto} S.,  2007, \mn@doi [\apj] {10.1086/521806}, \href
  {http://adsabs.harvard.edu/abs/2007ApJ...669..663M} {669, 663}

\bibitem[\protect\citeauthoryear{Mesinger, Furlanetto  \& Cen}{Mesinger
  et~al.}{2011}]{Mesinger2011}
Mesinger A.,  Furlanetto S.,   Cen R.,  2011, MNRAS, 411, 955

\bibitem[\protect\citeauthoryear{{Mo}, {Mao}  \& {White}}{{Mo}
  et~al.}{1998}]{Mo1998}
{Mo} H.~J.,  {Mao} S.,   {White} S.~D.~M.,  1998, \mn@doi [\mnras]
  {10.1046/j.1365-8711.1998.01227.x}, \href
  {http://adsabs.harvard.edu/abs/1998MNRAS.295..319M} {295, 319}

\bibitem[\protect\citeauthoryear{{Mutch}, {Geil}, {Poole}, {Angel}, {Duffy},
  {Mesinger}  \& {Wyithe}}{{Mutch} et~al.}{2016}]{Mutch2015}
{Mutch} S.~J.,  {Geil} P.~M.,  {Poole} G.~B.,  {Angel} P.~W.,  {Duffy} A.~R.,
  {Mesinger} A.,   {Wyithe} J.~S.~B.,  2016, \mn@doi [\mnras]
  {10.1093/mnras/stw1506}, \href
  {http://adsabs.harvard.edu/abs/2016MNRAS.462..250M} {462, 250}

\bibitem[\protect\citeauthoryear{{Oesch} et~al.,}{{Oesch}
  et~al.}{2016}]{Oesch2016}
{Oesch} P.~A.,  et~al., 2016, \mn@doi [\apj] {10.3847/0004-637X/819/2/129},
  \href {http://adsabs.harvard.edu/abs/2016ApJ...819..129O} {819, 129}

\bibitem[\protect\citeauthoryear{{Planck Collaboration}}{{Planck
  Collaboration}}{2015}]{Planck-Collaboration2015}
{Planck Collaboration} 2015, preprint, \href
  {http://adsabs.harvard.edu/abs/2015arXiv150201589P} {} (\mn@eprint {arXiv}
  {1502.01589})

\bibitem[\protect\citeauthoryear{{Poole}, {Angel}, {Mutch}, {Power}, {Duffy},
  {Geil}, {Mesinger}  \& {Wyithe}}{{Poole} et~al.}{2016}]{Poole2016}
{Poole} G.~B.,  {Angel} P.~W.,  {Mutch} S.~J.,  {Power} C.,  {Duffy} A.~R.,
  {Geil} P.~M.,  {Mesinger} A.,   {Wyithe} S.~B.,  2016, \mn@doi [\mnras]
  {10.1093/mnras/stw674}, \href
  {http://adsabs.harvard.edu/abs/2016MNRAS.459.3025P} {459, 3025}

\bibitem[\protect\citeauthoryear{{Powell}, {Bournaud}, {Chapon}  \&
  {Teyssier}}{{Powell} et~al.}{2013}]{Powell2013}
{Powell} L.~C.,  {Bournaud} F.,  {Chapon} D.,   {Teyssier} R.,  2013, \mn@doi
  [\mnras] {10.1093/mnras/stt1036}, \href
  {http://adsabs.harvard.edu/abs/2013MNRAS.434.1028P} {434, 1028}

\bibitem[\protect\citeauthoryear{{Song} et~al.,}{{Song}
  et~al.}{2016}]{Song2015}
{Song} M.,  et~al., 2016, \mn@doi [\apj] {10.3847/0004-637X/825/1/5}, \href
  {http://adsabs.harvard.edu/abs/2016ApJ...825....5S} {825, 5}

\bibitem[\protect\citeauthoryear{{Tacchella}, {Trenti}  \&
  {Carollo}}{{Tacchella} et~al.}{2013}]{Tacchella2013}
{Tacchella} S.,  {Trenti} M.,   {Carollo} C.~M.,  2013, \mn@doi [\apjl]
  {10.1088/2041-8205/768/2/L37}, \href
  {http://adsabs.harvard.edu/abs/2013ApJ...768L..37T} {768, L37}

\bibitem[\protect\citeauthoryear{{Trac}, {Cen}  \& {Mansfield}}{{Trac}
  et~al.}{2015}]{Trac2015}
{Trac} H.,  {Cen} R.,   {Mansfield} P.,  2015, \mn@doi [\apj]
  {10.1088/0004-637X/813/1/54}, \href
  {http://adsabs.harvard.edu/abs/2015ApJ...813...54T} {813, 54}

\bibitem[\protect\citeauthoryear{{Trenti}, {Santos}  \& {Stiavelli}}{{Trenti}
  et~al.}{2008}]{Trenti2008}
{Trenti} M.,  {Santos} M.~R.,   {Stiavelli} M.,  2008, \mn@doi [\apj]
  {10.1086/592037}, \href {http://adsabs.harvard.edu/abs/2008ApJ...687....1T}
  {687, 1}

\bibitem[\protect\citeauthoryear{{Trenti}, {Stiavelli}, {Bouwens}, {Oesch},
  {Shull}, {Illingworth}, {Bradley}  \& {Carollo}}{{Trenti}
  et~al.}{2010}]{Trenti2010}
{Trenti} M.,  {Stiavelli} M.,  {Bouwens} R.~J.,  {Oesch} P.,  {Shull} J.~M.,
  {Illingworth} G.~D.,  {Bradley} L.~D.,   {Carollo} C.~M.,  2010, \mn@doi
  [\apjl] {10.1088/2041-8205/714/2/L202}, \href
  {http://adsabs.harvard.edu/abs/2010ApJ...714L.202T} {714, L202}

\bibitem[\protect\citeauthoryear{{Trenti}, {Perna}  \& {Jimenez}}{{Trenti}
  et~al.}{2015}]{Trenti2015}
{Trenti} M.,  {Perna} R.,   {Jimenez} R.,  2015, \mn@doi [\apj]
  {10.1088/0004-637X/802/2/103}, \href
  {http://adsabs.harvard.edu/abs/2015ApJ...802..103T} {802, 103}

\bibitem[\protect\citeauthoryear{{Waters}, {Di Matteo}, {Feng}, {Wilkins}  \&
  {Croft}}{{Waters} et~al.}{2016a}]{Waters2016b}
{Waters} D.,  {Di Matteo} T.,  {Feng} Y.,  {Wilkins} S.~M.,   {Croft} R.~A.~C.,
   2016a, \mn@doi [\mnras] {10.1093/mnras/stw2000}, \href
  {http://adsabs.harvard.edu/abs/2016MNRAS.tmp.1188W} {}

\bibitem[\protect\citeauthoryear{{Waters}, {Wilkins}, {Di Matteo}, {Feng},
  {Croft}  \& {Nagai}}{{Waters} et~al.}{2016b}]{Waters2016}
{Waters} D.,  {Wilkins} S.~M.,  {Di Matteo} T.,  {Feng} Y.,  {Croft} R.,
  {Nagai} D.,  2016b, \mn@doi [\mnras] {10.1093/mnrasl/slw100}, \href
  {http://adsabs.harvard.edu/abs/2016MNRAS.461L..51W} {461, L51}

\makeatother
\end{thebibliography}







\bsp
\label{lastpage}
\end{document}